# ONSET OF AIR ENTRAINMENT BY A SMOOTH PLUNGING JET UNDER ATMOSPHERIC PRESSURE.

Alain Cartellier*, Juan Lasheras °

*LEGI (INPG-CNRS-UJF), BP53 38041 Grenoble, cedex 9 France, ° Department of Mechanical and Aerospace Engineering, University of California, La Jolla, California, 92093-0411.

**ABSTRACT**

The onset of air entrainment by a smooth vertical liquid jet impacting a pool of the same liquid has been experimentally determined. The ranges of parameters covered complement those considered by Lin & Donnelly (1969). The influence of the jet curvature is clarified. A model based on the viscous stress acting on the interface proves to be in good agreement with the experiments in the limit of very viscous liquids. For less viscous liquids, the critical capillary number happens to be mainly controlled by the ratio of the dynamic viscosity of the air and of the liquid. Besides, the observed evolution of the critical capillary number agrees with the mechanism suggested by Eggers (2001) that is based on the cusp dynamic.

## 1 - INTRODUCTION

Air entrainment is a common phenomenon encountered in nature (breaking waves, torrents…) as well as in industry. For processes, the objective could be either to optimize the exchange between phases such as in aeration devices, liquid jet pumps… or to diminish and possibly to eliminate air entrainment such as during the filling of containers, for coating operations, or for the design of bow waves. The importance of applications explains the large amount of literature dedicated to air entrainment. Yet, many fundamental questions related in particular with the critical conditions of air entrainment, the entrained gas flow rate, the resulting bubble size distribution, the depth of penetration of bubbles are not fully resolved. This situation is well illustrated by the review of Bin [1] in the case of a single jet plunging vertically into a pool of the same liquid. Hereafter, attention is focused on the critical conditions of air entrainment.

For turbulent jets, it has been early recognized that the jet roughness strongly affects the entrainment process [see for ex. 2-4]. Such a feature may partly explain the significant scatter of the experimental data as illustrated in [1] since the jet stability depends on many parameters including the injector design and the distance above the pool. Yet the observations are only partially explained by available models even if considerations related with the jet stability are incorporated in the analyses. The key point is that there is no definite agreement on the elementary mechanisms that must be accounted for. To quote a few, Van de Sande & Smith [2,3] and McKeogh & Ervine [4] ground their analysis on the boundary layer developing in the gas along the incoming jet. Lizzi and Prosperetti [5] and Bonetto, Drew and Lahey [6] consider the stability of the gas film that forms below the free surface. Zhu, Oguz and Prosperetti [7] et al. demonstrate how a liquid bulge interacting with the pool surface creates a cavity and successfully entraps air.

For smooth laminar jets, the absence of interface roughness should render the air entrainment problem more tractable. Yet, no truly predictive model is available for the onset of air entrainment. Existing proposals will be discussed in the last section. On the experimental side, data on smooth cylindrical jets have been mainly gathered by Lin & Donnelly [8] (see also Lin [9] and a few additional measurements by Perry [10] and Cummins [11]). Lin & Donnelly observed that, for a given jet radius, smooth jets of viscous liquids (the fluids they used are given in Table 1) do entrain air when the jet velocity at impact exceeds a critical value $V_c$. They established the following empirical criterion for the onset of air entrainment:

$$We = 10\ Re^{0.74} \quad (1)$$

where the Weber (We) and Reynolds (Re) numbers are based on the jet velocity and diameter at impact. Equation (1) has been found valid within about 10% (the largest deviations being +28% and -19%) for jet Reynolds numbers evaluated at the onset of air entrainment i.e. $Re=2V_cR/\nu$ in the range 8 to 1300. For fluids of lower viscosity (not shown in Table 1) and thus higher jet Reynolds numbers, the authors mention the apparition of irregularities at the jet surface that drastically modify the entrainment process so that eq.(1) no longer holds. Although eq.(1) is widely accepted, it raises various questions.

First, eq.(1) fails to provide any onset condition for a plane jet. It happens that the range of jet radii considered by Lin & Donnelly has been limited (the jet radius scaled by the capillary length ranged between 0.66 and 2.5). An investigation of the influence of the jet curvature is therefore worth undertaking.

Second, one may wonder whether the rupture in the We(Re) relationship for Re above 1300 corresponds to a drastic modification of the driving mechanisms, or if the corresponding experiments were affected by some defect and were not truly representative of a smooth jet. This question is also connected with an on-going debate concerning the ability of perfectly smooth jet to entrain air. Zhu et al. [7] argued that an interface disturbance is required for air entrainment to occur. Although Lin & Donnelly's experiments (as well as various coating-type experiments) seem to contradict this statement, one may possibly invoke the presence of tiny interfacial disturbances that remained undetected during the experiments. Eggers [12] also concluded that for liquids of low viscosity, no cusp can be formed and therefore that air entrainment by a smooth water jet is impossible.

Third, the eq.(1) does not involve any parameter related with the upper phase. There are growing evidences that the upper fluid plays a role even if it is gaseous. The experiments by Robertson et al. [13] indicate that the critical velocity should increase by lowering the gas pressure. In a study related with coating, Simpkins & Kuck [14] demonstrated how an adjustment of the gas pressure can delay the onset of air entrainment. In addition, Linan & Lasheras [15] established a model of the gas shroud formed when air entrainment is present, in which the gas dynamic viscosity plays a central role. More recently, by combining the cusp solution exhibited by Jeong & Moffatt [16] with the gas flow, Eggers [12] shows that the cusp equilibrium can be destroyed by the movement of the gas phase.

These questions lead us to revisit the problem of the onset of air entrainment (OAE) for smooth cylindrical round jets. The experiments are described in the next section. The results are presented in last section where the mechanisms are also discussed.

Table 1. Physical properties of the liquids used in refs [1-2].

| Reference | $\mu$ (kg/m s) | $\rho$ (kg/m$^3$) | $\sigma$ (N/m) | a (mm) |
|---|---|---|---|---|
| Lin - 400 | 3.62E-01 | 1245 | 0.061 | 2.23 |
| Lin - 300 | 2.96E-01 | 1242 | 0.059 | 2.27 |
| Lin - 200 | 2.06E-01 | 1236 | 0.062 | 2.26 |
| Lin -100 | 9.90E-02 | 1223 | 0.062 | 2.27 |
| Lin - 130 | 1.28E-01 | 876 | 0.0305 | 1.88 |
| Lin -50 | 5.00E-02 | 1207 | 0.061 | 2.27 |
| Lin -25 | 2.50E-02 | 1189 | 0.058 | 2.23 |
| Lin -10 | 9.60E-03 | 1157 | 0.058 | 2.26 |

## EXPERIMENTS

### Idealized situation

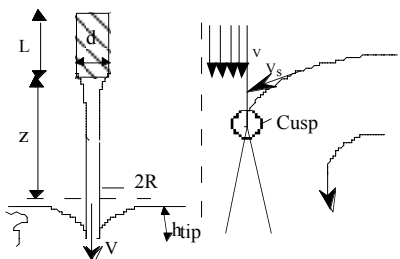

Fig.1: Sketch of a plunging jet.

Ideally, the situation investigated consists of a perfectly smooth vertical and stable round liquid jet impacting a pool of the same liquid in a given gas ambiance (Fig.1). If the velocity profile at impact is fixed, nor the detail of injector design (diameter d, length L, internal design…) nor its height z above the free surface enter the problem. The critical velocity $V_c$ corresponding to the onset of air entrainment should then depend only on the jet diameter 2R at impact plus the fluids densities and viscosities, their surface tension $\sigma$ and the acceleration of gravity (Fig.1). Two natural scales appear, one for the length $a^2=\sigma/(\rho g)$ and one for the velocity $U=\sigma/\mu$ (both based here on liquid properties). The unknown $V_c$ can be expressed as a capillary number $Ca=V/U=\mu V/\sigma$ at the critical conditions, namely $Ca_{crit}=V_c/U$. The controlling parameters are the density $\rho_G/\rho$ ratio, the viscosity $\mu_G/\mu$ ratio, the jet radius measured in terms of the capillary length R/a, and a Reynolds number based on the above natural scales $Re_{cap}=Ua/\nu$. Hence one expects:

$Ca_{crit} = f (a/R, \mu_G/\mu, \rho_G/\rho, Re_{cap}, \text{jet velocity profile})$ (2)

$Re_{cap}$ is related with the Morton number $Mo=g\mu^4/\rho\sigma^3$ by $Re_{cap}^2 Mo=1$, and as such, it characterizes the liquid properties (Morton numbers are also of common use in the literature related with film coating). In addition, $Re_{cap}$ provides a measure of the importance of gravity since it also writes $Re_{cap}=U/(ga)^{1/2}$.

Experiments have been dedicated to span at least a decade in terms of R/a and to extend the range of fluids properties. As we shall see, the jet Reynolds numbers Re at onset has increased up to about a few thousands. A special attention has been paid to investigate the influence of the jet velocity profile. The only parameters which were kept fixed are the gas properties.

### Experimental set-up and procedure

To vary the control parameters, different fluids have been considered including silicon oils, water-glycerol mixtures, perfluorocarbon (Table 2) as well as de-ionized and filtered water. The gas was ambient air at atmospheric pressure for all the tests. The injectors were also varied, both in size and type. The inner diameter at exit d ranged from 0.2 up to 6mm. Two tapered injectors were carefully designed to provide uniform velocity profiles (diameter contraction ratio of 5, no straight section at exit). The others injectors were needles with a straight length L varying from 1.2 to 500 diameters in order to obtain flat and parabolic velocity profiles. To minimize pressure fluctuations and jet disturbances, the flow was ensured by means a vessel pressurized with regulated air. It passed through a battery of high accuracy flow meters equipped with needle valves and calibrated for every fluid.

Table 2. Liquids physical properties.

| Reference | $\mu$ (kg/m s) | $\rho$ (kg/m$^3$) | $\sigma$ (N/m) | a (mm) |
|---|---|---|---|---|
| Si - 100 | 8.42E-02 | 926 | 0.019 | 1.46 |
| Si - 50 | 4.85E-02 | 948 | 0.019 | 1.44 |
| Canola Oil | 6.59E-02 | 913.5 | 0.029 | 1.79 |
| Si - 20 | 1.79E-02 | 941 | 0.020 | 1.46 |
| Si - 10 | 8.80E-03 | 916.5 | 0.019 | 1.47 |
| Si - 5 | 4.66E-03 | 913 | 0.019 | 1.45 |
| Si - 1 | 8.94E-04 | 812 | 0.018 | 1.53 |
| Si - 0.65 | 5.01E-04 | 754.5 | 0.018 | 1.55 |
| Mixture 2 | 6.69E-03 | 1121 | 0.054 | 2.22 |
| Mixture 3 | 4.67E-03 | 1107 | 0.054 | 2.23 |
| FC-77 | 1.53E-03 | 1776.6 | 0.018 | 1.01 |

The jet impinged the free surface in the center of a receiving tank of square section. To maintain a fixed free-surface level, the liquid in the pool overflows along two opposite sides of the tank. This design provides an undisturbed optical access to the jet at the impact. The other pair of walls were extended above the free surface in order to visualize the inverted meniscus without any optical distortion. Images were collected with CCD cameras equipped with a macro zoom and using parallel backlightning to optimize the interface detection. The spatial resolution was between 3 and 7µm/pixel both in air and in liquids (the magnification was adapted to the jet size). The interface position was deduced from the grey level distribution using the *NIH image 1.60 software*. Across an edge, the grey level evolves from 0 to 255 within 3-4 pixels. Hence, the resulting relative uncertainty on the jet diameter never exceeded a few percent. The jet velocity at impact V was deduced from the liquid flow rate and from the measured jet radius R at impact assuming a flat velocity profile. The typical relative uncertainty on velocity was 10%.

To determine the onset of air entrainment, experiments have been conducted according to the same procedure. For a fixed height z, the liquid flow rate was very progressively increased until *continuous* air entrainment occurred. For this condition, images of the jet in the air were then collected to measure the actual jet diameter at impact and at onset. After every modification of the flow rate, some delay was allowed to stabilize the velocity fields both in the jet and in the receiving tank. Experiments were systematically repeated, and the reproducibility was found to be of the order of the uncertainty mentioned above.

## Observations

Although the process by which air entrainment by a smooth, stable jet occurs is described in the literature, it is worthwhile to clarify some of its features. When a smooth viscous jet impacts a pool, a collar-like meniscus forms. At very low velocities, the meniscus lays above the undisturbed free-surface and the surface velocity in the vicinity of the impact is centrifugal. Increasing the jet velocity induces a decrease of the meniscus height and ultimately leads to an inverted meniscus as illustrated Fig.2-top). In that case, the liquid velocity in the vicinity of the interface becomes centripetal. Increasing further the jet velocity, the depth and the lateral extend of this dip (or gas trumpet) increase up to a point where air is suddenly entrained. Just at onset, a thin gas film forms around the liquid jet that plunges far below the free-surface (Fig.3); bubbles detach from this compound jet and are entrained in the pool. For liquids of large viscosity, the OAE coincides with the formation of this long gas film.

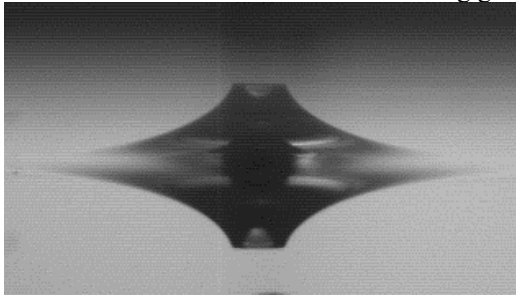

Fig.2: Inverted meniscus formed below the free surface before air entrainment for Canola oil (top, with a mirror image due to the free surface) and for Si-10 (bottom).

At lower liquid viscosity, say below about $10^{-2}$ kg/ms, a trumpet also forms but its depth below the free surface is significantly smaller than for viscous liquids. In addition, the gas collar looses its symmetry. Instead, a pointed tip appears, as shown Fig.2-bottom, from which the first bubbles detach. This is reminiscent of the instabilities observed in coating [17].

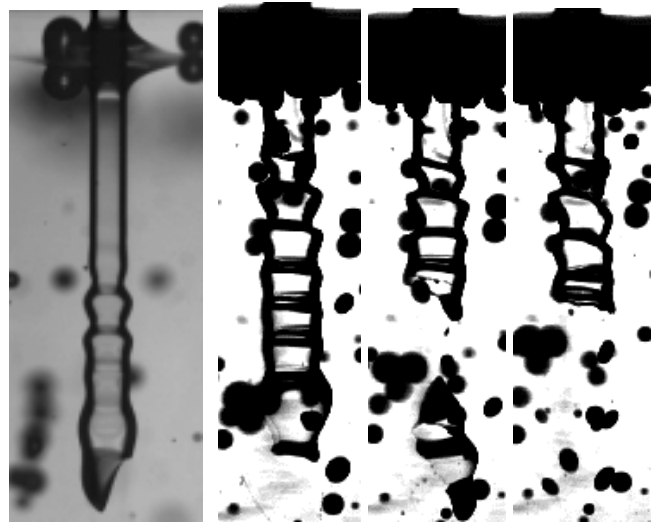

Fig.3: Structure of the gas film below the free surface after onset. The sequence to the right illustrates the periodic rupture of this structure and the formation of bubbles. Tiny bubbles also arise from rim instability (second image from the left).

## Sensitivity to flow conditions

Although the transition to air entrainment is abrupt, various precautions must be fulfilled in order to obtain meaningful measurements of OAE for laminar jets.

First, the pool must be free of bubbles since even a single bubble trapped in the vicinity of the jet impact region significantly delays the OAE. Similarly, disturbances of the pool free surface modify the OAE. This problem arises mainly for high jet Reynolds numbers (let say above 800) which correspond to high liquid flow rates and thus to a stronger liquid circulation in the receiving pool. The use of a wider tank allows to get rid of this effect. It has been checked that $V_c$ is independent of the tank size for depths between 30 to 120 injector diameters, and side lengths over d between 15 and 120. Another phenomena worth to be mentioned is the spurious apparition of a vortex. This mode of air entrainment has been observed only for jet Reynolds numbers above a few hundreds (hence for low viscosity fluids). It was avoided by using large tanks provided that initial conditions are still enough. All the data presented have been gathered in absence of vortex.

Second, any jet instability must be avoided. Instead of quantifying interface distortions, the evolution of the OAE with the height z of the injector above the pool has been systematically studied. For small to moderate z/d, $V_c$ happens to be constant within the reproducibility of the measurements. Examples of such evolutions are given for three typical sets corresponding to low (Fig.4), moderate (Fig.5) and high (Fig.6) jet Reynolds numbers. Above some distance, which depends on the injector design, instabilities have enough time to develop and $V_c$ decreases sharply as expected in presence of interfacial roughness. Beyond this point, the air entrainment process becomes unsteady and accordingly the reproducibility of measurements gets poorer.

Additionally, for high Re, $V_c$ was found to increase at very short distances from the nozzle due to incomplete relaxation of the velocity profile and jet contraction. By avoiding both the developing region and the unstable zone, it happens that the critical velocity is weakly dependent (even independent if one considers the experimental uncertainty) of the injector design and varies only with the jet velocity and size at impact (for a given fluid). Thus, the collected OAE data truly correspond to stable smooth jets.

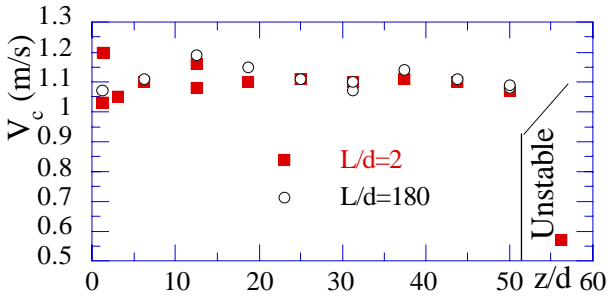

Fig.4: Evolution of the critical jet velocity $V_c$ with the impact distance at Re~30 and for two injectors (fluid Si-50).

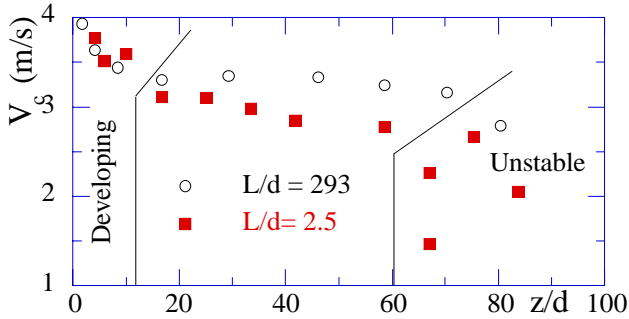

Fig.5: Evolution of the critical jet velocity $V_c$ with the impact distance at Re~300 and for two injectors (fluid Si-10).

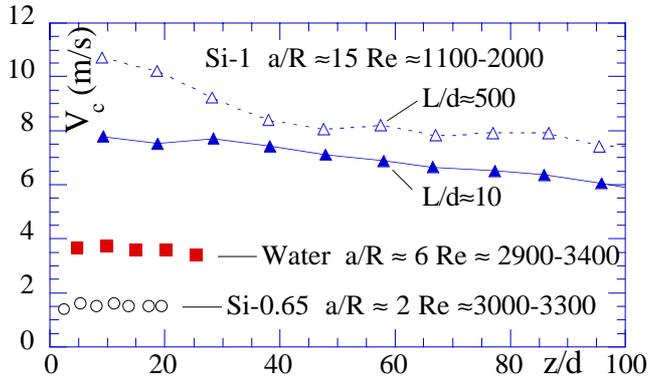

Fig.6: Evolution of the critical jet velocity $V_c$ with the impact distance and for two injectors at large Re.

**Influence of the velocity profile**

The influence of the velocity profile on the OAE transition has been carefully tested by varying the injector design, and in particular, by changing the ratio L/d for capillary tubes. The critical velocity is higher for a parabolic profile (high L/d) than for a flat profile (L/d =O(1)). However, at small Re, say below 100, this increase is negligible (Fig.4). At Re about a few hundreds, it remains comparable with the reproducibility (for example, the increase is about 0.4m/s at Re=300 - see Fig.5). For the highest Re numbers considered, the sensitivity to the velocity profile is more marked (up to 0.8-1.2m/s - Fig.6), but the dispersion on the critical velocity remains without acceptable limits (typically 15%). Therefore, flat and parabolic profiles are not longer distinguished, and the velocity profile parameter can be eliminated from eq.(2).

**CRITICAL CONDITIONS**

The full range of experimental conditions considered is given Fig.7 in terms of a/R and jet Reynolds number evaluated at onset. Re ranges from 5 to about 5000 while a/R spans more than a decade. The magnitude of the critical velocity ranges typically from 1 to 10m/s.

The principal trends observed on viscous fluids by Lin & Donnelly, namely a strong decrease of the critical velocity with the liquid viscosity and a moderate decrease with the jet radius, are shown here to remain valid over an extended range of fluids. In particular these trends hold down to a liquid viscosity as small as $10^{-3}$ kg/ms. One can also notice in Fig.7 that both the Silicon oil Si-10 and the Canola oil fall on the same curve. Their physical properties are different but their $Re_{cap}$ (or Morton numbers) are nearly identical. Hence, referring to eq.(2), the role of the gas may not be so crucial in the viscous limit.

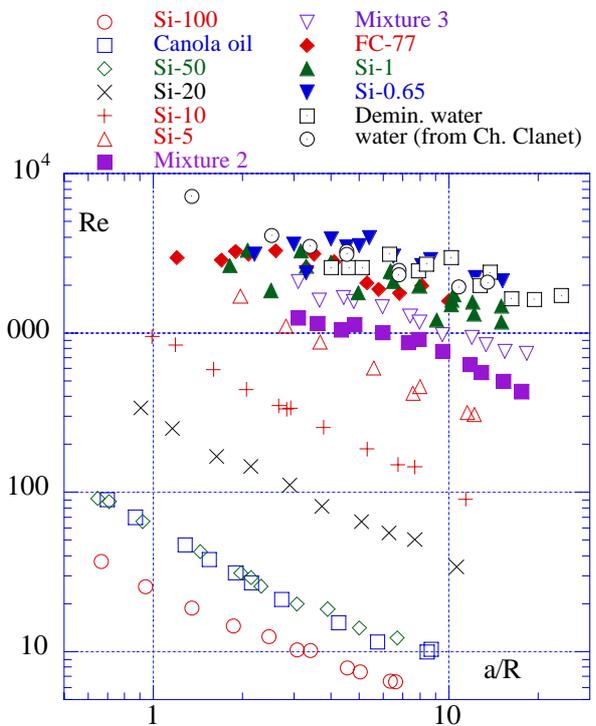

Fig.7: Evolution of the jet Reynolds number at onset with the jet curvature for various fluids. The second set of data for water are from Ch. Clanet (private communication). From bottom to top, the curves correspond to increasing $Re_{cap}$.

Reporting the data on a We(Re) plot, it is clear that eq.(1) does not hold beyond Re of order 1000 as argued by Lin & Donnelly (Fig.8.). As stated in the introduction, their conclusion was based on the observation of interfacial disturbances that appear for low viscosity fluids in their experiments. No jet disturbances have been observed here (let us note in passing that the magnitude of $V_c$ is far too small to trigger any Taylor type instability - exploited notably in diesel engines - so that surface disturbances are indeed unlikely to occur). Hence, it is concluded that the loss of validity of eq.(1) does exists for smooth jets. This rupture probably implies a qualitative change in the mechanism driving the onset of air entrainment.

Let us now turn toward the dimensionless parameters entering eq.(2). The critical Capillary number, which is the only one involving $V_c$ is plotted versus a/R for various liquids in Fig.9. Two sets of data from Lin & Donnelly, that cover a significant range in terms of a/R, and a set of data for water due to Ch. Clanet have been added to complete the plot. $Ca_{crit}$ happens to be fairly large, of the order of 10, for the more viscous fluids considered, indicating that the viscous stresses in the liquid are important in that regime. As the liquid viscosity diminishes, so does $Ca_{crit}$ that decreases down to about O(0.01) for water. It is also clear from Fig.9 that the influence of the jet curvature depends on the fluid properties and hence on the flow regime.

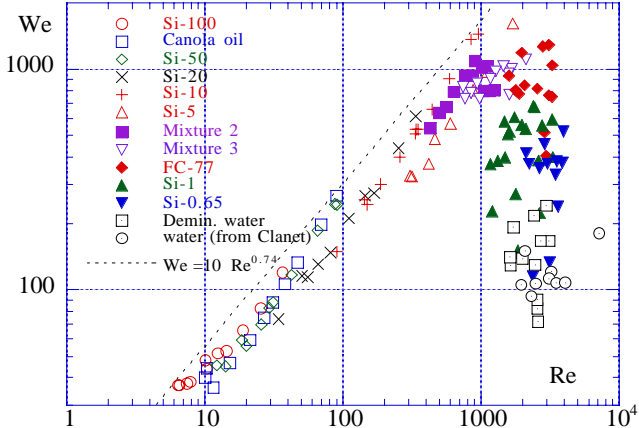

Fig.8: We versus Re at onset (Note that a/R is not fixed. Instead it covers the range shown Fig.7).

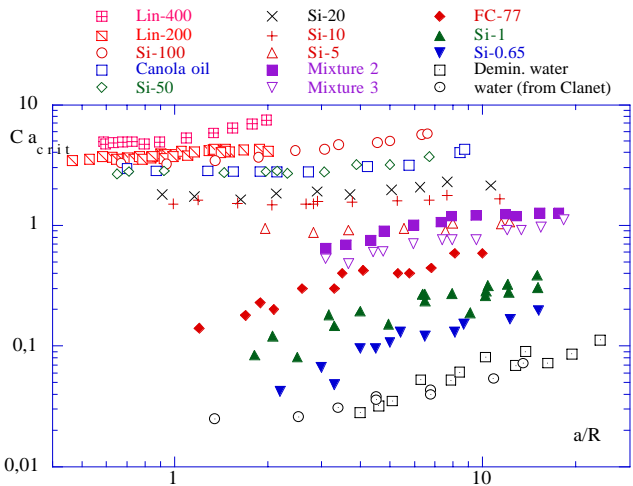

Fig.9: Critical capillary number versus a/R for various fluids. From top to bottom, the curves correspond to increasing $Re_{cap}$.

**Role of the jet curvature**

For a smooth stable jet, the key to the onset of air entrainment is the free surface deformation. The "gas trumpet" profile is controlled by the stress balance along the normal to the interface:

$$- P_L + \tau_L + P_G - \tau_G = \sigma \kappa \qquad (3)$$

where the pressures and the normal stresses are evaluated in each phase and in the vicinity of the interface. The magnitudes of the various stresses are set by the entrained velocity field generated in the pool and by the movement of the gas phase. Whatever the response of the gas phase, the starting point is the existence of a downward traction due to the entrainment of the outer liquid by the incoming jet. Roughly, this traction can originate from a depression in the vicinity of the jet impact point: this is a possible driving mechanism in the limit of large jet Reynolds numbers. For low jet Reynolds number, the pressure distribution in the bulk of the fluid is nearly hydrostatic, and the traction originates from the viscous stresses. Whatever the physical origin of the traction, its magnitude increases with the jet velocity. In response to this solicitation, the free surface deforms as a negative meniscus, and it plunges deeper below the undisturbed level. The evolution of the depth of penetration is illustrated Fig.10 in the limit of low Re number. In the same time, the profile in a meridian plane becomes steeper, and this process holds until the interface become almost vertical at the tip. Increasing further the traction, no room is left for the interface to deform, and the system assumes another equilibrium shape with the formation of a gas film along the jet [15]. The mere presence of the gas film uncouples the jet from the outer fluid. As the film extends further below the surface, the mixing layer starts at a larger depth so that the upper part of the meniscus corresponds almost to hydrostatic conditions. The depth of penetration of this gas film is controlled by extra mechanisms involving instabilities and possibly gravity (see Fig.3). The important point here is that the interface near its tip is almost vertical at onset. This statement is also valid in presence of a cusp [16]. Consequently, the curvature in a meridian plane tends toward zero while the radius of curvature in the second principal plane tends to the jet radius. Owing to the limited depth of penetration, and since the jet remains cylindrical, this radius is well approximated by the jet radius R measured at the location of the undisturbed free-surface. Thus, at onset, the total curvature $\kappa$ tends to 1/R at the tip. In other words, the radius curvature increases the pressure in the liquid by an amount $\sigma/R$ at the tip. This extra pressure opposes the interface deformation, so that an extra stress and therefore an higher jet velocity are required to overcame the effect of the jet curvature. This conclusion is consistent with the observed trends shown Fig.9. It also indicates that the jet curvature should appear as a correction of the critical conditions for a plane jet.

The above reasoning remains valid when the movement in the gas is taken into account. Indeed, the gas flow can be approximated as a lubrication layer that tends to widen the gas channel. As the gas flow is (partly) set by the boundary conditions on the jet and the "trumpet" interfaces, its characteristic velocity also scales with V and so does the gas pressure. Hence, even if the gas phase assumes at some point a driven role in the distortion of the interface, the stresses it generate remain proportional to the jet velocity. The conclusion concerning the role of the jet curvature is therefore the same.

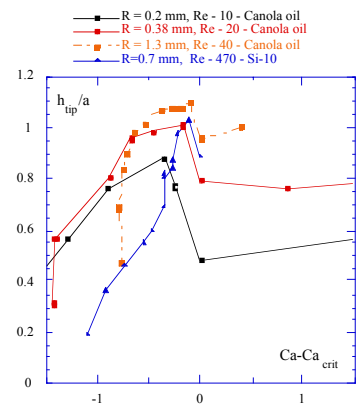

Fig.10: Evolution of the meniscus depth with Ca. Beyond the onset (Ca>$Ca_{crit}$), $h_{tip}$ represents the depth of the point where the gas structure diameter is minimum (the later increases further below due to friction with the outer liquid).

## Tentative model in the viscous limit

In the limit of very small jet Reynolds numbers, the experiments presented above indicate that the viscous traction in the liquid is a key ingredient for the onset of air entrainment. We therefore tentatively proposed a model on that basis without considering the movement in the gas phase nor the role of the cusp that probably forms at the extremity of the gas trumpet [18]. The point is to evaluate the terms entering eq.(3) at the tip of the meniscus and at onset. We have seen in the previous paragraph that $\kappa$ tends to $1/R$ in these conditions. For very small Re, the pressure field in the liquid is nearly hydrostatic, so that the excess pressure at the tip is evaluated as $\rho g h_{tip}$, and the depth of penetration $h_{tip}$ at onset is of the order of the capillary length as shown Fig.10.

For the normal viscous stress $\tau_L = \mu \, dV_n/dn$, we note that tangential velocity $V_s$ along the interface increases from zero at a distance of about a few capillary lengths away from the jet up to a value of the order of V at the tip (Fig.1). As shown by Schiffer et al. [19], such a stretching of the interface strongly deforms the meniscus compared with the static solution. The free surface being a streamline, $dV_n/dn$ equals $dV_s/ds$ where s is the length along the interface. However, the tangential velocity distribution along the interface $V_s(s)$ is not known. Moreover, from the few measurements of $V_s$ we have done, $V_s$ happens to vary most rapidly in the vicinity of the tip and thus the largest stress arises in the tip region. There is therefore no obvious scaling for $dV_s/ds$ (unless cusp models, which exhibit also a rapid change in $V_s$, are considered - see the discussion at the end of the paper). To proceed, let us go back to $\tau_L = \mu \, dV_n/dn$ and let us consider the diffusion of momentum from the accelerating free surface boundary. This situation is analogous to the transient momentum diffusion away from a boundary with a velocity increasing with time. Using the capillary length as a measure of the interface extension, the diffusion time is about $a/V_s \sim a/V$, so that the diffusion thickness is of order $(\nu a/V)^{1/2}$ and thus $\tau_L \sim \rho \nu^{1/2} V^{3/2} a^{-1/2}$. Collecting the above estimates, the condition of equilibrium (3) at the tip and at onset writes:

$$\rho g h_{tip} + \sigma / R = \tau_L \sim \rho \nu^{1/2} V^{3/2} a^{-1/2} \qquad (4)$$

Eq.(4) expresses the balance between the viscous stress generated along the surface and the over pressure to be overcome for OAE to occur. The latter is the sum of an hydrostatic contribution and a capillary pressure due to the finite size of the jet. In terms of the dimensionless numbers introduced above, the relation (4) transforms into:

$$Ca_{crit} \sim Re_{cap}^{-1/3} (h_{tip}/a + a/R)^{2/3} \qquad (5)$$

where $h_{tip}/a$ is O(1).

Contrary to the empirical formula (1), eq.(5) also applies to planar (2D) jets. In that case, a/R tends to zero and the corresponding critical velocity results from a balance between viscous traction and gravity. No data is available to test directly the validity of eq.(5) on 2D jets. Instead, the experimental obtained on cylindrical jets were exploited to evaluate $Ca_{crit}$ for fixed values of a/R. The results, including data from Lin & Donnelly, are shown Fig.11. The predicted scaling $Ca_{crit} \sim Re_{cap}^{-1/3}$ for a given a/R ratio is in good agreement with the experiments. This holds for $Re_{cap}$ less than about 20 that correspond to jet Reynolds numbers below 130. This indicates a posteriori that the influence of surfactants (if any since we used clean fluids) is not crucial in this problem. Also, the precise shape of the interface (cusped or not) at the tip has not been considered. Yet it cannot be concluded that it is unimportant since the gas viscosity does enter eq.(5) although it is expected to play a role. In terms of Morton number, $Ca_{crit} \sim M^{1/6}$. Let us note that similar trends with weak exponents have been reported for the onset of air entrainment in coating systems.

Concerning the evolution of $Ca_{crit}$ with a/R, one expects from eq.(5) that the quantity $Ca_{crit}^{3/2} Re_{cap}^{1/2}$ linearly growths with a/R. As shown Fig.12, the experiments agree with this trend although the range of radii investigated has been limited. An interesting point to notice is the variation of the slope for the different liquids. This may well reflect the influence of the parameter $\mu_G/\mu$ which is absent from eq.(5). Let us also point out that the various curves have almost the same origin, so that the depths of the meniscus profile before onset have indeed similar magnitudes for all these fluids.

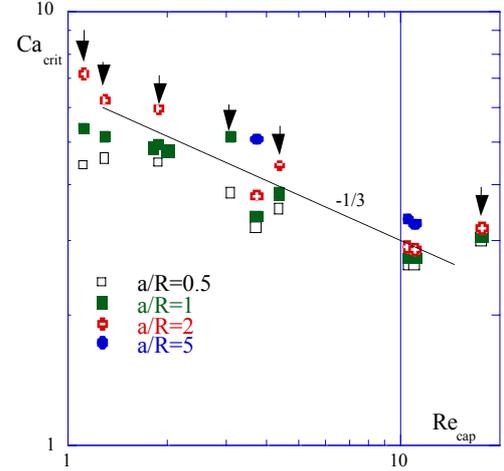

Fig.11: Test of the scaling predicted by eq.(5) in the low Re regime (arrows indicate Lin & Donnelly's data).

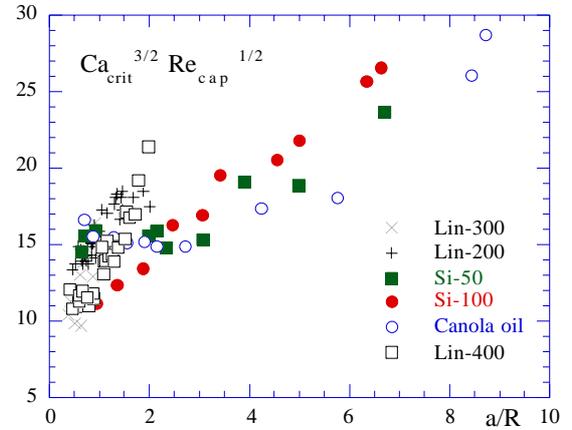

Fig.12: Test of the scaling $Ca_{crit}^{3/2} Re_{cap}^{1/2} \sim a / R$ expected from eq.(5) in the limit of small Re.

## Behaviors outside the viscous limit

Eq.(5) fails to reproduce the observations for fluid of viscosity below about $10^{-1}$ kg/ms and hence for jet Reynolds numbers above a hundred. Instead, as shown Fig.13, $Ca_{crit}$ monotonously decreases with $Re_{cap}$ but it exhibits different regimes, the slope in a log-log plot increasing from -1/3 in the viscous limit to -1/4 at intermediate $Re_{cap}$ and then to -3/4 for very large $Re_{cap}$. The rupture occurring at $Re_{cap}$ about $10^4$ is the same as the one identified in the We(Re) plot of Fig.8.

In the very high jet Reynolds number limit, one may be tempted to neglect the viscous stresses and o estimate the depression due to the entrained flow field as $\rho V^2$. The corresponding onset condition would become:

$$V^2 / (\rho g a) \sim h_{tip}/a + a / R \qquad (6)$$

Such a scaling fails to reproduce the experiments (it would correspond to $Ca_{crit} \sim Re_{cap}^{-1/2}$ in Fig.13). In particular, it gives too much credit to gravity. From experiments, Ca decreases from about 10 to about 0.1 while $Re_{cap}=U/(ga)^{1/2}$ increases from unity up to $10^5$. Hence, the Froude number $V_c/(ga)^{1/2}$ is most of the time very high ($10^3$-$10^4$) so that gravity is negligible. Only in the viscous limit does the Froude number decreases down to about 10, and that explains the relative success of eq.(5) which is based on a meniscus not deviating too much from the static solution.

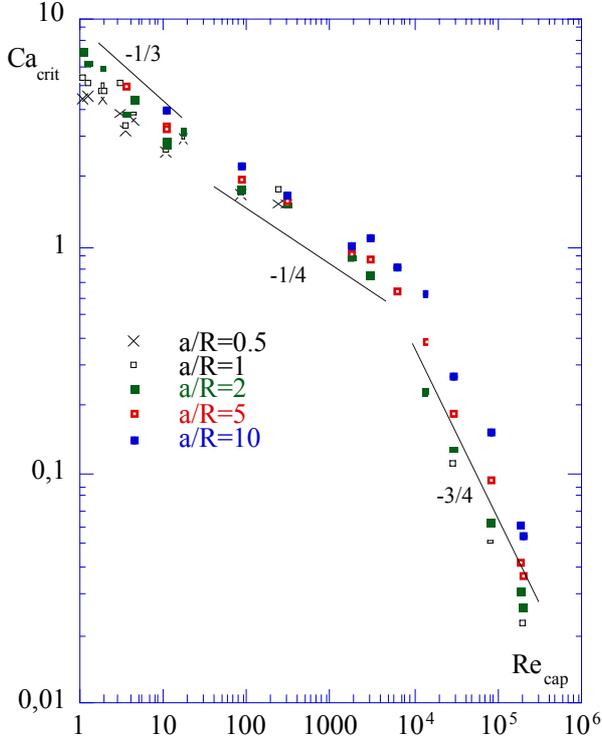

Fig.13. Evolution of $Ca_{crit}$ with $Re_{cap}$ for various a/R ratio.

Going back to large Re, and discarding gravity, $Re_{cap}$ disappears from the list of parameter in eq.(2). In addition, the gas inertia is negligible compared with surface tension (the Weber number for the gas is at most unity in our experiments) so that effect of the density ratio can also be discarded. Therefore eq.(2) reduces to:

$$Ca_{crit} = f(a/R, \mu_G/\mu) \qquad (7)$$

The effect of $\mu_G/\mu$ enters the problem via the stresses generated in the gas. As already mentioned, one expects a built up of the lubrication pressure in the thin air gap. Its magnitude depends on the air gap thickness. The probable presence of a cusp enters the problem at this point since its affects the length scale to be considered. Indeed, Jeong & Moffatt [16] have found a family of solutions for the liquid flow around a cusp that holds at any capillary number. They found that the shape of the interface obeys a universal form of the type $\eta(x) = c (x/X)^{3/2} + \sqrt{2} (x R_{cusp})^{1/2}$ where x denotes the vertical distance to the tip. At large distance from the tip, the scaling $\eta(x) = c (x/X)^{3/2}$ involves some outer distance X that depends on the way the flow is generated. For the plunging jet, X may be taken as the jet radius, and the constant c reflects this outer flow. Very near the tip, the outer length scale becomes unimportant, and the scaling $\eta(x) \propto (x R_{cusp})^{1/2}$ depends solely on the radius of curvature at the tip $R_{cusp}$. Jeong & Moffatt have also shown that $R_{cusp}$ exponentially decreases with a capillary number based on the liquid viscosity and a velocity scale connected with the outer flow. Namely:

$$R_{cusp} \propto \exp(-c\, Ca) \qquad (8)$$

Based on this scale, and following Eggers [12], the normal stresses are expected to be of order $\mu_G V R_{cusp}^{-1}$ in the gas phase and of order $\mu V R_{cusp}^{-1/4} X^{-3/4}$ in the liquid phase. Air entrainment arises when the equilibrium in eq.(1) is destroyed so that one expects the rupture when the radius of curvature at the cusp becomes of the order of $\mu_G/\mu^{-4/3}$. Hence, for a plane jet, one should have:

$$Ca_{crit} \propto \log(\mu/\mu_G) \qquad (9)$$

To test this prediction, all the data presented Fig.13 have been reported as $Ca_{crit}$ versus $\mu/\mu_G$ for given values of a/R (see Fig.14). It happens that the scaling predicted by eq.(9) is in good agreement with the experiments. In particular, the log trend holds for different a/R ratio, confirming thus that the jet curvature intervenes as a correction. Some discrepancies can be noticed at low $\mu/\mu_G$ values. In particular, the data for water, that correspond to the smallest $Ca_{crit}$, are neatly below those obtained from Silicon oils (Si-1 and Si-0.65). The later neatly departs from the log trend. As surfactants should not play a key role here, the origin of such a difference is unclear.

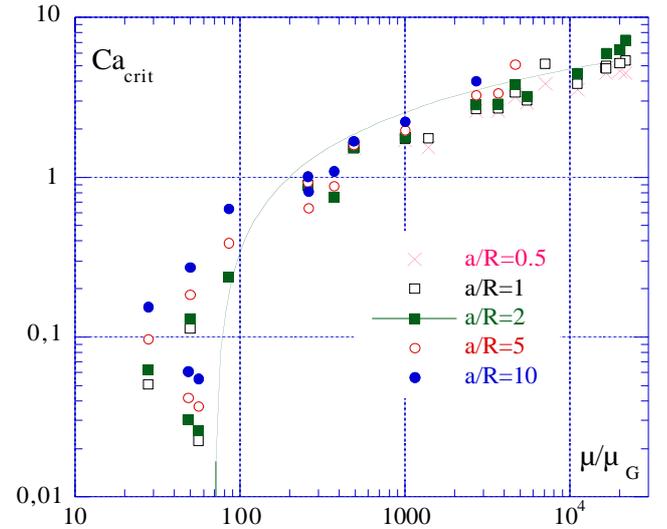

Fig.14: Evolution of $Ca_{crit}$ with the dynamic viscosity ratio for various values of a/R. The continuous line is a log best fit.

The destabilizing role of the gas phase anticipated by Eggers appears to be the key mechanism for the onset of air entrainment by smooth jets. Aside incorporating the jet curvature those role has been clarified, it is likely that the onset would be modified by the apparitions of instabilities at the gas rim. Those may possibly explain the departures to the log law that appear Fig.14. In the viscous limit, the log trend seems well established so that the cusp seems to be an essential ingredient even in that limit. It is difficult to reconcile this statement with the relative success of the viscous limit model that discards any cusp influence. Possibly, the viscous model holds as an the asymptotic behavior of air entrainment in the limit of for large $\mu/\mu_G$. This point deserves to be clarified.

## CONCLUSION

The experiments presented in this paper have clearly shown that air entrainment occurs for a smooth jet, that is even in absence of interfacial disturbances. The jet curvature has been shown to contribute only as an extra resisting stress. A model has been proposed in the limit of very viscous jets, which explains how the critical capillary number varies with the liquid properties and the jet size. To generalize the approach to less viscous fluids, it has been shown that gravity is much less important but that the movement in the gas phase has to be taken into account. In that respect, the proposal of Eggers concerning the key role of the cusp that forms at the tip of the miniscus has been shown valid. Also, the main parameters affecting the critical capillary number being the dynamic viscosity ratio $\mu/\mu_G$ and the jet radius, the rupture identified on a Weber-Reynolds plot by Lin & Donnelly disappears. Instead, various regimes can be distinguished which can be tentatively classified as a very viscous regime, a regular cusp regime at intermediate viscosity and, for an even lower viscosity, a regime implying saw-tooth instabilities of the gas rim such as the one shown Fig.2.

## NOMENCLATURE

a capillary length = $(\sigma/\rho g)^{1/2}$ (m)
c constant (-)
d injector diameter at exit (m)
g gravity (m/s$^2$)
L injector length (m)
n direction normal to the interface
P pressure (N/m$^2$)
R jet radius at impact (m)
$R_{tip}$ cusp radius of curvature (m)
s length along the interface (m)
V jet velocity at impact (m/s)
$V_c$ critical jet velocity at impact (m/s)
$V_s$ velocity tangent to the interface (m/s)
$V_n$ velocity normal to the interface (m/s)
X length scale (m)
x distance to the tip (m)
z injector height above the surface (m)

$\mu_G$ gas dynamic viscosity (kg/m s)
$\mu$ liquid dynamic viscosity (kg/m s)
$\nu_G$ gas cinematic viscosity (m$^2$/s)
$\nu$ liquid cinematic viscosity (m$^2$/s)
η interface profile
κ interface curvature (m$^{-1}$)
$\rho_G$ gas density (kg/m$^3$)
ρ liquid density (kg/m$^3$)
σ surface tension (N/m)
τ viscous stress (N/m$^2$)